\documentclass[journal]{IEEEtran}

\ifCLASSINFOpdf
\else
\fi

\usepackage{setspace}%????
\usepackage{bbm}
\usepackage[cmex10]{amsmath}
\usepackage{amssymb}
\usepackage{cite}
\usepackage{graphicx}
\usepackage{array,color}
\usepackage{amsmath}
\allowdisplaybreaks
\usepackage{stfloats}
\usepackage{graphicx}
\usepackage{epstopdf}
\usepackage{subfigure}
\usepackage{tabularx}
\usepackage{epsfig,epsf,color,balance,cite}
\usepackage{algorithmic}
\usepackage{algorithm}
\usepackage{url}
\usepackage{bm}
\usepackage{multirow}

\usepackage{amsthm}
\ifodd 0
\usepackage{soul}
\usepackage{color}
\setstcolor{red}

\newcommand{\rev}[1]{{\color{red}#1}} %revise of the text
\newcommand{\del}[1]{\st{#1}} %deleting the text

\newcommand{\com}[1]{\textbf{\color{red} (COMMENT: #1)}} %comment of the text
\newcommand{\response}[1]{\textbf{\color{green} (RESPONSE: #1)}} %response to comment
\else

\newcommand{\rev}[1]{#1}

\newcommand{\del}[1]{}

\newcommand{\com}[1]{}
\newcommand{\comg}[1]{}
\newcommand{\response}[1]{}
\fi
\title{\vspace{-0.3cm}\huge {Fast Channel Estimation for IRS-Assisted OFDM}}

\author{ \vspace{-0.2cm}
	Beixiong Zheng,~\IEEEmembership{Member,~IEEE}, Changsheng You,~\IEEEmembership{Member,~IEEE}, 
	and Rui Zhang,~\IEEEmembership{Fellow,~IEEE} 
	\vspace{-0.9cm}
	\thanks{\vspace{-0.35cm}
		
		The authors are with the Department of Electrical and Computer Engineering, National University of Singapore, Singapore 117583,
		email: \{elezbe, eleyouc, elezhang\}@nus.edu.sg.
		\vspace{-0.8cm}

	}
}

\begin{document}
%\markboth{Submitted to IEEE Transations on XXX}{SKM: My IEEE article}
\maketitle
%\vspace{-1cm}
\begin{abstract}
In this letter, we study efficient channel estimation for an intelligent reflecting surface (IRS)-assisted orthogonal frequency division multiplexing (OFDM) system to achieve minimum training time. 
First, a fast channel estimation scheme with reduced OFDM symbol duration is proposed for arbitrary frequency-selective fading channels. Next, under the typical condition that the IRS-user channel is line-of-sight (LoS) dominant, another fast channel estimation scheme based on \rev{the novel concept of \emph{sampling-wise IRS reflection variation} is proposed.} Moreover, the pilot signal and IRS training reflection pattern are jointly optimized for both proposed schemes. Finally, the proposed schemes are compared in terms of training time and channel estimation performance via simulations, as well as against benchmark schemes.
	
\end{abstract}
%\vspace{-0.1cm}
\IEEEpeerreviewmaketitle

\vspace{-0.1cm}
\section{Introduction}
%\IEEEPARstart{I}{ntelligent}
%Leveraging the recent advances in metamaterial technology, 
%intelligent reflecting surface (IRS) and its various equivalents have recently emerged as a promising cost-effective
%solution to improve the wireless communication spectrum and energy efficiency

As an enabling technology for smart and reconfigurable wireless communication environment, intelligent reflecting surface (IRS) has recently drawn a great deal of attention. By leveraging a large number of low-cost passive elements that are able to reflect signals with adjustable amplitudes and/or phase shifts, \rev{IRS is capable of significantly enhancing the wireless communication system throughput in an energy-efficient and cost-effective manner \cite{qingqing2019towards,Huang2019Reconfigurable}.}
However, the promising gains brought by IRS critically depend on the channel state information (CSI) that is practically difficult to acquire, due to the large number of channel coefficients associated with massive IRS reflecting elements.
%, which is in sharp contrast to  conventional wireless systems with active access point (BS) and user only.
This issue becomes more challenging for orthogonal frequency division multiplexing (OFDM) systems with frequency-selective fading channels, which incur even more channel coefficients due to the multi-path delay spread. 
\rev{Some prior works \cite{yang2019intelligent,zheng2019intelligent,zheng2020intelligent} have addressed this problem for IRS-assisted OFDM systems by estimating the cascaded IRS channels via different IRS training reflection designs (e.g., the ON/OFF-based design \cite{yang2019intelligent} and the discrete Fourier transform (DFT) matrix-based design \cite{zheng2019intelligent,zheng2020intelligent}).
Moreover, a novel element-grouping strategy was proposed by properly grouping adjacent IRS elements into a sub-surface, which provides a flexible system trade-off between training overhead/design complexity and passive beamforming performance by adjusting
the size of each sub-surface \cite{yang2019intelligent,zheng2019intelligent}. }
However, in the above works as well as others for IRS channel estimation under flat-fading channels (see, e.g., \cite{jensen2019optimal,chen2019channel,you2019progressive}), the number of (OFDM) pilot symbols should be no less than the number of all links including the direct link
and the cascaded IRS links (whose number equals to the number of IRS reflecting elements/sub-surfaces), which can still be prohibitively high and thus is inapplicable to practical systems with insufficient pilot symbols/training time.  
%wang2019channel,jensen2019optimal,chen2019channel

% Moreover, most of the existing works \cite{zheng2019intelligent,zheng2020intelligent,yang2019intelligent,jensen2019optimal,wang2019channel,you2019progressive,chen2019channel} focus on the IRS reflection pattern design for IRS channel estimation, while the pilot sequence design is simply treated or completely missing. Note that to achieve optimal channel estimation performance over limited training time and power, the pilot sequence should be jointly designed with the IRS reflection pattern, which, however, is not fully investigated yet, to the best of our knowledge.
%since the number of OFDM sub-carriers is typically much larger than the maximum number of delayed paths in practical systems, there exists great redundancy for channel estimation

Motivated by the above, this letter studies more efficient channel estimation design for an IRS-assisted OFDM system to achieve minimum training time.  
%the long OFDM symbol duration
%Specifically, we propose two efficient channel estimation schemes for different channel setups.
First, for arbitrary frequency-selective fading channels, we propose a fast channel estimation scheme by shortening the duration of OFDM symbols, which achieves much less training time as compared to the schemes given in  \cite{zheng2019intelligent,zheng2020intelligent,yang2019intelligent}.
Next, under the typical scenario with single-path IRS-user channel (or line-of-sight (LoS) dominant channel in practice) due to their nearby deployment, we propose another fast channel estimation scheme based on \rev{the novel concept of \emph{sampling-wise IRS reflection variation}, which creates artificial linear and time-variant (ALTV) cascaded channels within one OFDM symbol to
 facilitate fast IRS channel estimation. It is shown that the latter proposed scheme in general achieves much better channel estimation performance with less training time as compared to the former one, both with their corresponding jointly optimized IRS training reflection pattern and pilot signal design.}
%Numerical results are provided to  compare the two proposed schemes in terms of minimum mean squared error (MSE) of the estimated channels, as well as against benchmark schemes.

\vspace{-0.1cm}
\section{System Model and Problem Description}\label{sys}
\begin{figure}[!t]
	\centering
	\includegraphics[width=2.0in]{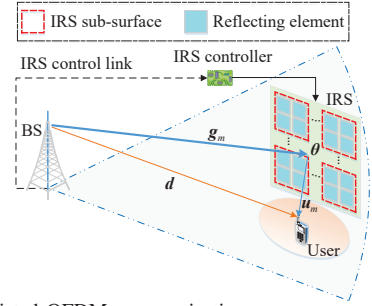}
	\caption{IRS-assisted OFDM communication.}
	\label{system}
	\vspace{-0.5cm}
\end{figure}
% composed of $K$ reflecting elements 
As shown in Fig. \ref{system},
we consider the basic IRS-assisted point-to-point communication system, where an IRS 
is deployed to assist the transmission from a base station (BS) to a
user, both of which
are equipped with a single antenna.\footnote{\rev{The proposed channel estimation can be applied to the multi-user downlink communication where the users estimate their individual channels in parallel as well as the uplink communication with multi-antenna BS by treating each BS antenna/user as an equivalent user/BS antenna in the downlink case. }\vspace{-0.8cm}}
%By adopting the elements-grouping strategy 
As in \cite{zheng2019intelligent,zheng2020intelligent,yang2019intelligent}, the IRS composed of ${M}_0$ reflecting elements
is divided into $M$ sub-surfaces, denoted by the set ${\cal M} \triangleq \{1,2,\ldots, M\}$, each consisting of $\mu= {M}_0/M$ (assumed to be an integer) adjacent elements that share a common reflection coefficient for reducing the implementation complexity.
%Note that given the channel coherence time, 
%the number of sub-surfaces $M$ (or equivalently the size of sub-surface $\mu$) is adjustable to
% provide a flexible performance trade-off between
%the channel estimation overhead and IRS reflection performance.
Moreover, the IRS is connected to a smart controller that adjusts its reflection coefficients and exchanges information with the BS via a separate reliable wireless link \cite{qingqing2019towards}.
%Note that the IRS is practically composed of a large number of passive reflecting elements to maximize its refection power, 
%which, however, incurs high overhead/complexity for channel estimation and reflection optimization.
%By grouping adjacent elements of the IRS with high channel correlation into a sub-surface to share a common reflection coefficient \cite{yang2019intelligent}, the complexity of channel estimation and reflection design can be significantly reduced.
%, thus simplifying the system design and saving the deployment cost.
%To simplify the system design and reduce the deployment cost, we assume that multiple adjacent elements form a sub-surface to share a common phase shift \cite{yang2019intelligent} so that the complexity in terms of channel estimation and reflection design can be significantly reduced. 
%Accordingly, the IRS composed of $K$ reflecting elements is divided into $M$ sub-surfaces, each of which consists of ${K}=K/M$ adjacent elements, e.g., ${K}=4$ as illustrated in Fig.~\ref{system}.
%Moreover, the IRS is connected to a smart controller to enable dynamic adjustment of its elements' individual reflections.
In this letter, the quasi-static block fading channel model is assumed for all the channels involved, which remain approximately constant within the channel coherence time.

Let ${{\bm d}}\in {\mathbb{C}^{L_d\times 1}}$, ${{\bm g}}_m \in {\mathbb{C}^{L_1\times 1 }}$, and 
${{\bm u}}_{m} \in {\mathbb{C}^{L_2\times 1 }}$ 
 denote the baseband
equivalent channels in the time domain for the BS$\rightarrow$user, BS$\rightarrow$sub-surface $m$, and sub-surface $m$$\rightarrow$user links, respectively, where $L_d$, $L_1$, and $L_2$ denote the maximum multi-path delay spreads (normalized by the sampling period $1/B$ with $B$ denoting the system bandwidth) of these links.	 
Let ${\bm \theta}\triangleq[{\theta_1},  {\theta_2},\ldots,{\theta_M}]^T=[ e^{j \phi_1},  e^{j \phi_2},\ldots, e^{j \phi_M}]^T$ denote the equivalent reflection coefficients of IRS sub-surfaces, where $\phi_m \in [0, 2\pi)$ represents the phase shift of sub-surface $m$ and the reflection amplitudes of all elements are set to one or the maximum value for simplicity.
%To maximize the signal power reflected by the IRS and reduce the hardware cost, we set $\beta_m=1, \forall  m\in {\cal M} $ and only consider the phase-shift design of the IRS.
Thus, the effective cascaded channel from 
the BS to the user via sub-surface $m$ can be expressed as the convolution of the BS$\rightarrow$IRS channel, the IRS reflection coefficient,
and the IRS$\rightarrow$user channel, which is given by
%\vspace{-0.1cm}
\begin{align}\label{conv_ch}
{{\bm g}}_m \ast \theta_m \ast {{\bm u}}_{m}  =\theta_m {{\bm g}}_m \ast  {{\bm u}}_{m}
=\theta_m {{\bm q}}_{m}
\end{align}
where ${{\bm q}}_{m} \triangleq {{\bm g}}_m \ast  {{\bm u}}_{m} \in {\mathbb{C}^{L_r\times 1 }} $ denotes the cascaded BS$\rightarrow$IRS$\rightarrow$user channel (without IRS phase shifts) associated with sub-surface $m$ and $L_r=L_1+L_2-1$ is the maximum delay spread of the cascaded BS$\rightarrow$IRS$\rightarrow$user channel.
\rev{Furthermore, we take the maximum delay spread of the direct BS$\rightarrow$user channel and the cascaded BS$\rightarrow$IRS$\rightarrow$user channel, i.e., $L=\max \{L_d,L_r\}$ and apply zero-padding \cite{zheng2020intelligent}.}

%Let $L=\max\{L_r, L_d \}$ denote the maximum delay spread of the effective time-domain channel between the users and BS, while letting ${\bm q}_{m}$ and ${\bm d}$ denote the 
%zero-padded cascaded BS$\rightarrow$IRS$\rightarrow$user (reflecting) channel of ${{\bm q}}_{m}$ and zero-padded BS$\rightarrow$user (direct) channel of ${{\bm d}}$ for user $k$, with the zero padding lengths of $L-L_r$ and $L- L_d$, respectively.
As a result, the superimposed channel impulse response (CIR) from the BS to the user by
combining the direct BS$\rightarrow$user channel and the cascaded BS$\rightarrow$IRS$\rightarrow$user channel, denoted by ${\bm h} \in {\mathbb{C}^{L\times 1 }}$, is given by
%\vspace{-0.1cm}
\begin{align}\label{superposed}
{\bm h}={\bm d}+{\bm Q} {\bm \theta}
\end{align}
where ${\bm Q}=[{\bm q}_{1},{\bm q}_{2},\ldots,{\bm q}_{M}] \in {\mathbb{C}^{L\times M }}$ denotes
the cascaded BS$\rightarrow$IRS$\rightarrow$user channel matrix by stacking ${\bm q}_{m}$ with $m=1,\ldots,M$.
%where ${\bm \Theta}=\text{diag} \left( {\bm \theta} \right)$ represents the diagonal reflection matrix of the IRS.
According to \eqref{superposed}, it is required to estimate 
the direct channel ${\bm d}$ and
the cascaded channels $\{{\bm q}\}_{m=1}^M$ for the IRS-assisted OFDM system.\footnote{Specifically, the user sends back its estimated CSI to the BS, which designs the IRS reflection coefficients for data transmission and sends them to the IRS controller for implementation.\vspace{-0.2cm}}  Thus, the total number of channel
coefficients to be estimated is $L(M+1)$.
%for the passive beamforming design of ${\bm \theta}$

%With OFDM, the total bandwidth $B$ is equally divided into $N$ sub-carriers. 

Consider the OFDM transmission with $N$ sub-carriers, where each OFDM symbol of length $N$ is appended by a cyclic prefix (CP) of length $L_{cp}\ge L$ to mitigate the inter-symbol interference and we usually have $N \gg L$ in practice.
In the existing literature on IRS channel estimation (see, e.g., \cite{zheng2019intelligent,zheng2020intelligent,yang2019intelligent,jensen2019optimal,chen2019channel,you2019progressive}), at least $M+1$ pilot symbols are required to estimate the total $L(M+1)$ channel coefficients.
As a result, it requires $\eta_0=(M+1)(N+L_{cp})$ sampling periods for channel training in the case of OFDM, which is practically high when $M$ is large.
To reduce channel training time for the IRS-assisted OFDM, we present two new fast channel estimation
schemes by exploiting the fact that $N \gg L$ under two different channel setups in the following two sections, respectively.
% by removing the great sampling redundancy due to $N \gg L$
%, and derive
%the fundamental limits of these schemes on the minimum training overhead and the maximum
%number of supportable users in the IRS-assisted multi-user OFDMA system

\vspace{-0.1cm}
\section{Channel Estimation Based on Short-OFDM-Symbol-Wise Varying IRS Reflection}\label{Sch1}
%--------------------------------------------------------------------
\newcounter{TempEqCnt}                        % 创建临时变量TempEqCnt
\setcounter{TempEqCnt}{\value{equation}} % 将当前公式序号 赋给TempEqCnt
\setcounter{equation}{7}                           % 当前公式序号变为x，x等于长公式应有的序号减1.
\begin{figure*}
	\vspace{-0.4cm}
	\begin{equation}\label{rec_sub-surface}
	\resizebox{0.9\textwidth}{!}{$\hspace{-0.6cm}
		\underbrace{\begin{bmatrix}
			r_{m,0}\\
			r_{m,1}  \\
			\vdots \\
			r_{m,N-1}
			\end{bmatrix}}_{{\bm{{r}}}_m}
		\hspace{-0.1cm}=\hspace{-0.1cm} \begin{bmatrix}
		\theta_m^{(0)}q_{m,0} &0  &\cdots &0   &\theta_m^{(0)}q_{m,L-1} &\cdots &\theta_m^{(0)}q_{m,1}\\
		\theta_m^{(1)}q_{m,1} & \theta_m^{(1)}q_{m,0} &     &0    	&0  		&     &\theta_m^{(1)}q_{m,2}\\
		\vdots   		& \vdots  		&       &\vdots   		&\vdots    		&     	&\vdots\\
		\theta_m^{(L-1)}q_{m,L-1}  	&\theta_m^{(L-1)} q_{m,L-2}  &       & 0       		&0          	&      	&0\\
		0        		&\theta_m^{(L)}q_{m,L-1}&       & 0       		&0         		&       &0\\
		\vdots   		& \vdots  		&     	&\vdots   		&\vdots    		&       &\vdots\\
		0        		&0        		& \cdots& \theta_m^{(N-1)}q_{m,L-1} &\theta_m^{(N-1)}q_{m,L-2}  &\cdots &\theta_m^{(N-1)}q_{m,0}
		\end{bmatrix}\hspace{-0.15cm}
		\underbrace{\begin{bmatrix}
			x_0\\
			x_1  \\
			\vdots \\
			x_{N-1}
			\end{bmatrix}}_{{\bm{{ x}}}}
		\hspace{-0.1cm}= \hspace{-0.1cm}
		\underbrace{\begin{bmatrix}
			\theta_m^{(0)}&&\\
			&\ddots &\\
			&&\theta_m^{(N-1)}
			\end{bmatrix}}_{{\bm{{ \Theta}}_m}}\hspace{-0.15cm}
		\underbrace{\begin{bmatrix}
			x_0 	&x_{N-1} 		&\cdots 	&x_{N-L+1}     \\
			x_1 	&x_0 		&\cdots 	&x_{N-L+2}     \\
			x_2 	&x_1 		&\cdots 	&x_{N-L+3}     \\
			\vdots  & \vdots     &\ddots  	&\vdots   \\
			x_{L-1} 	&x_{L-2} 	&\cdots 	&x_0\\
			\vdots  & \vdots  	    &   	&\vdots   \\
			x_{N-1} 	&x_{N-2} 	&\cdots 	&x_{N-L}
			\end{bmatrix}}_{{\bm{{X}}}}\hspace{-0.15cm}
		\underbrace{\begin{bmatrix}
			q_{m,0}\\
			q_{m,1} \\
			\vdots \\
			q_{m,L-1}
			\end{bmatrix}}_{{\bm q}_{m}}\hspace{-0.3cm}
		$
	}\hspace{-0.2cm}
	\vspace{-0.2cm}
	\end{equation}
	\hrulefill
	\vspace{-0.6cm}
\end{figure*}
\setcounter{equation}{\value{TempEqCnt}}
%--------------------------------------------------------------------
%Since the number of sub-carriers in the OFDM-based broadband communications is
%typically much larger than that of delayed propagation paths, there exists a great redundancy for channel estimation
%As pointed out in \cite{zheng2019intelligent,zheng2020intelligent}, the 
%\section{Channel Estimation}
In this section, we propose a new fast channel estimation scheme (referred to as Scheme 1) for arbitrary frequency-selective fading channels.
%, by shortening the duration of each time-domain OFDM symbol.
Specifically, the length of each OFDM symbol (without CP) is shortened to $N_0$ with $N \gg N_0\ge L$ for channel training.
% to ensure the resolvability of each channel link of $L$ delayed propagation paths
Let ${\bm s}\triangleq  \left[S_{0},S_{1},\ldots, S_{N_0-1}  \right]^T$ denote the short-OFDM symbol in the frequency domain, which is first transformed into the time domain via an $N_0$-point inverse DFT (IDFT) and then appended by a CP of length $L_{cp}\ge L$.
%where $i=1,\ldots,I_0$ with $I_0$ denoting the number of short-OFDM symbols for channel training.
%Let ${\bm x}_k^{(i)}\triangleq  \left[x_{N_0-1}^{(i)},\ldots, x_{N_0-1}^{(i)},x_{0}^{(i)},x_{1}^{(i)},\ldots, x_{N_0-1}^{(i)}  \right]^T$ denote the $i$-th short-OFDM symbol transmitted by the BS, where $i=1,\ldots,I_0$ with $I_0$ denoting the number of short-OFDM symbols for channel training.
After CP removal at the user side, the $i$-th received short-OFDM symbol in the frequency domain, denoted by ${\bm z}^{(i)}$, can be expressed as
%\vspace{-0.1cm}
\begin{align}\label{receive}
	{\bm z}^{(i)}&=  {\bm S}{\bm F} \left({\bm d}+{\bm Q} {\bm \theta}^{(i)}\right)+ {\bm v}^{(i)}
\end{align}
where 
$i=1,\ldots,I_0$ with $I_0$ denoting the number of short-OFDM symbols for channel training,
%${\bm y}^{(i)}\triangleq \left[Y^{(i)}_{0},Y^{(i)}_{1},\ldots, Y^{(i)}_{N-1}   \right]^T$ is the $i$-th received short-OFDM symbol,
${\bm S} =\text{diag} \left({\bm s}\right)$ is the diagonal matrix of each short-OFDM symbol ${\bm s}$,
${\bm F}$ is an $N_0 \times L$ matrix consisting of the first $L$ columns of the $N_0 \times N_0$ unitary DFT matrix,
${\bm \theta}^{(i)}$ denotes the IRS training reflection coefficients during the transmission of the $i$-th short-OFDM symbol,
and ${\bm v}^{(i)}\sim {\mathcal N_c }({\bm 0}, \sigma^2{\bm I}_{N_0} )$ is the additive white Gaussian noise (AWGN) vector at the user with $\sigma^2$ being the noise power.
%and ${\bm v}^{(i)}\triangleq \left[V_{0}^{(i)},V_{1}^{(i)},\ldots,V_{N-1}^{(i)}   \right]^T \sim {\mathcal N_c }({\bm 0}, \sigma^2{\bm I}_N )$ is the additive white Gaussian noise (AWGN) vector at the BS with $\sigma^2$ being the noise power.
By denoting ${\tilde{\bm S}}={\bm S}{\bm F}$,
${\tilde{\bm Q}}=\left[{\bm d}, {\bm Q} \right]$, and $ {\tilde{\bm \theta} }^{(i)}= \begin{bmatrix}1\\{\bm \theta}^{(i)}\end{bmatrix}$, \eqref{receive} can be written in a compact form as ${\bm z}^{(i)}={\tilde{\bm S}}{\tilde{\bm Q}}{\tilde{\bm \theta} }^{(i)}+{\bm v}^{(i)}, i=1,\ldots,I_0$. By stacking the received signal vectors $\left\{{\bm z}^{(i)}\right\}_{i=1}^{I_0}$ into ${\bm Z}=[{\bm z}^{(1)},{\bm z}^{(2)},\ldots,{\bm z}^{(I_0)}]$, we obtain
%\vspace{-0.1cm}
\begin{align}\label{receive_mat}
{\bm Z}={\tilde{\bm S}}{\tilde{\bm Q}} {\bm \Psi}+{\bm V}
\end{align}
where 
${\bm \Psi} \triangleq [{\tilde{\bm \theta} }^{(1)},{\tilde{\bm \theta} }^{(2)},\ldots,{\tilde{\bm \theta} }^{(I_0)}]$ denotes the IRS training reflection pattern matrix that collects all training reflection coefficients $\{{\tilde{\bm \theta} }^{(i)}\}_{i=1}^{I_0}$, and ${\bm V}=[{\bm v}^{(1)},{\bm v}^{(2)},\ldots,{\bm v}^{(I_0)}]$ denotes the corresponding AWGN matrix.
Based on \eqref{receive_mat}, we obtain the least-squares (LS) estimates of ${\bm d}$ and ${\bm Q}$ as
\begin{align}\label{LS_est0}
\left[{\hat{\bm d}}, {\hat{\bm Q}} \right]={\tilde{\bm S}}^\dagger{\bm Z}{\bm \Psi}^\dagger={\tilde{\bm Q}}+{\tilde{\bm S}}^\dagger{\bm V}{\bm \Psi}^\dagger
\end{align}
where ${\tilde{\bm S}}^\dagger=\left({\tilde{\bm S}}^H{\tilde{\bm S}}\right)^{-1}{\tilde{\bm S}}^H$ and ${\bm \Psi}^\dagger={\bm \Psi}^H \left({\bm \Psi}{\bm \Psi}^H\right)^{-1} $.
Note that for the LS channel estimation in \eqref{LS_est0}, $I_0\ge M+1 $ is required to ensure the existence of ${\bm \Psi}^\dagger$, \rev{implying that at least $M+1$ short-OFDM symbols are required to successfully estimate/distinguish both the direct channel ${\bm d}$ and the cascaded channel ${\bm Q}$ associated with the $M$ sub-surfaces.}
As such, accounting for the CP, it requires $\eta_1=(M+1)(N_0+L_{cp})$ sampling periods for the channel estimation based on the above short-OFDM-symbol-wise varying IRS training reflections.
\rev{It is noted that due to $N_0\ll N$, we have $\eta_1\ll\eta_0$, thus significantly reducing the training time, as compared to the schemes in \cite{yang2019intelligent,zheng2019intelligent,zheng2020intelligent}.}
Finally, the matrix inversion operation for computing ${\bm \Psi}^\dagger$ and ${\tilde{\bm S}}^\dagger$ in general has a cubic time complexity in terms of $M+1$ and $L$, respectively, and may lead to suboptimal channel estimation due to the potential noise enhancement if either ${\bm \Psi}$ or ${\tilde{\bm S}}$ is ill-conditioned, which thus requires a proper joint design of ${\bm \Psi}$ and ${\tilde{\bm S}}$.

The minimum mean squared error (MSE) of the LS channel estimation in \eqref{LS_est0} is given by
%\vspace{-0.1cm}
\begin{align}\label{MSE_cp}
\varepsilon_1&=\frac{1}{L(M+1)} {\mathbb E}\left\{  \left\|\left[{\hat{\bm d}}, {\hat{\bm Q}} \right]
-\left[{\bm d}, {\bm Q} \right]\right\|^{2}_F
\right\}
\notag \\
&=\frac{1}{L(M+1)}  {\mathbb E}\left\{ \Big\| {\tilde{\bm S}}^\dagger{\bm V}{\bm \Psi}^\dagger \Big\|^{2}_F\right\}\notag\\
&=\frac{1}{L(M+1)}  \text{tr}\left\{ \left({\bm \Psi}^\dagger \right)^H 
{\mathbb E}\left\{ {\bm V}^H ({\tilde{\bm S}}^\dagger) ^H {\tilde{\bm S}}^\dagger      {\bm V}   \right\}
{\bm \Psi}^\dagger	\right\}\notag\\
&\stackrel{(a)}{=}\frac{\sigma^2}{L(M+1)} \text{tr}\left\{\left({\bm F}^H {\bm S}^H {\bm S}{\bm F}\right)^{-1}\right\}
\text{tr}\left\{ \left({\bm \Psi}{\bm \Psi}^H\right)^{-1}	\right\}
%\vspace{-0.1cm}
\end{align}
where $(a)$ holds since ${\mathbb E}\left\{ {\bm V}^H ({\tilde{\bm S}}^\dagger) ^H {\tilde{\bm S}}^\dagger      {\bm V}   \right\} 
=\sigma^2 \text{tr}\left\{\left({\tilde{\bm S}}^H {\tilde{\bm S}}\right)^{-1}\right\} {\bm I}_{I_0}
=\sigma^2 \text{tr}\left\{\left({\bm F}^H {\bm S}^H {\bm S}{\bm F}\right)^{-1}\right\} {\bm I}_{I_0}$.
\rev{As such, the MSE minimization problem in \eqref{MSE_cp} can be equivalently decoupled into two sub-problems for minimizing $\text{tr}\left\{\left({\bm F}^H {\bm S}^H {\bm S}{\bm F}\right)^{-1}\right\}$ and
$\text{tr}\left\{ \left({\bm \Psi}{\bm \Psi}^H\right)^{-1}	\right\}$, respectively,} whose individual optimal values can be respectively achieved if and only if
${\bm S}^H {\bm S}=\gamma_1 {\bm I}_{N_0} $ with $\gamma_1$ being the average sub-carrier power and ${\bm \Psi}  {\bm \Psi}^H =(M+1){\bm I}_{M+1}$
according to \cite{kay1993fundamentals}.
\rev{This indicates that 
the optimal short-OFDM symbol should be equipowered over all sub-carriers (e.g., ${\bm s}=\gamma_1 {\bm 1}_{N_0 \times 1}$) and
the optimal IRS training reflection pattern ${\bm \Psi}$ should be an orthogonal matrix with each entry satisfying the unit-modulus constraint} (e.g., setting the IRS training reflection pattern ${\bm \Psi}$ as the $(M+1)\times (M+1)$ DFT matrix with each coefficient given by $\theta_m^{(i)}=e^{-j\frac{2\pi m (i-1) }{M+1}}$ with $ m=1,\ldots,M$ and $i=1,\ldots,M+1$).
Accordingly, we can obtain the minimum MSE of \eqref{MSE_cp} as $\varepsilon_1^{\min}=\frac{\sigma^2}{\gamma_1(M+1)} $, and have ${\tilde{\bm S}}^\dagger=\frac{1}{\gamma_1} {\tilde{\bm S}}^H$ and ${\bm \Psi}^\dagger=\frac{1}{(M+1)}{\bm \Psi}^H$ without the need of matrix inversion operation, thus avoiding the high (cubic-order) complexity.
%Accordingly, one optimal training design can be the 
%use the $(M+1)\times (M+1)$ DFT matrix as the reflection pattern ${\bm \Psi}$ with each IRS reflection coefficient given by
%\vspace{-0.4cm}
%\begin{align}\label{design_IRS}
%\theta_m^{(i)}=e^{-j\frac{2\pi m (t-1) }{M+1}}, \quad \forall m\in {\cal M}, \forall t \in {\cal T}
%\end{align}
\vspace{-0.4cm}
\section{Channel Estimation Based on Sampling-Wise Varying IRS Reflection}\label{Sampling}
%For the OFDM-based broadband communication system, the total bandwidth $B$ is equally divided into $N$ sub-carriers, which are indexed by $n\in {\cal N} \triangleq \{0,1,\ldots, N-1\}$. 
%For simplicity, we assume that the total transmission power at the user $P_t$ is equally allocated over the $N$ sub-carriers with the power at each sub-carrier given by $p_n=P_t/N,~\forall n \in {\cal N}$.
%To facilitate the joint resource allocation and IRS reflection at the BS, we consider a time-division duplexing (TDD) protocol for the uplink channel training over $\tau$ consecutive OFDM symbols during the time slots $n \in {\cal T}\triangleq \{1,2,\ldots,\tau\}$ of each channel coherence  time and assume the channel reciprocity for the CSI acquisition in the downlink communication based on the uplink training at the BS.
%Since the IRS elements have no transmit/receive RF chains, 
In this section, we consider the typical scenario with single-path IRS$\rightarrow$user channel (i.e., $L_2=1$) due to their practically short distance and propose another fast channel estimation scheme (referred to as Scheme 2) \rev{without changing the conventional OFDM symbol duration/structure.}

% yet achieving even less training time as compared to the previously introduced scheme based on short-OFDM symbols (to be specified later)
%\begin{align}\label{assumption}
%N \ge L(M+1)
%\end{align}
For the purpose of exposition, we consider the channel estimation at the user with only one OFDM pilot symbol by assuming $N \ge L(M+1)$ in the rest of this letter.\footnote{For the general case with arbitrary $L$ and $M$, the minimum number of OFDM pilot symbols for this scheme to estimate all the channel coefficients is $\left\lceil\frac{L(M+1) }{N} \right\rceil$, where $\left\lceil\cdot \right\rceil$ denotes the ceiling function.}
As such, the training time for Scheme 2 in terms of sampling periods is $\eta_2=N+L_{cp}$.
%To avoid inter-user interference and simplify the training design, we consider the disjoint pilot tone allocations for all the users in this paper,
%where each sub-carrier at each time slot is allocated to at most one user.
%Specifically, let $\delta_{k,n}$ indicate whether sub-carrier $n$ is allocated to user $k$ at time slot $n$, i.e., $\delta_{k,n}=1$ if sub-carrier $n$ is assigned to user $k$ at time slot $n$, and $\delta_{k,n}=0 $ otherwise. 
%Thus, we have $\delta_{k,n} \in\{0,1\}$ and $\sum_{k=1}^K \delta_{k,n} \le 1, \forall t \in {\cal T},  \forall n \in {\cal N}$.
%Here we denote ${\cal J}_{k}$ as the index set of the pilot tones assigned to user $k$ at time slot $n$, which is given by ${\cal J}_{k}\triangleq \left\{n | \delta_{k,n}=1 \right\}$.
%As the CSI is unknown \emph{a priori}, we consider the equal transmit power allocation for each user over the assigned $|{\cal J}_{k}|$ sub-carriers at each time slot $n$, where the transmit power of user $k$ on each assigned sub-carrier is given by $P/|{\cal J}_{k}|, \forall k\in {\cal K}, \forall t \in {\cal T}$.
%Moreover, due to
%the sum transmit power constraint at each user over the $N$ sub-carriers at
%each time slot $n$, we have $\sum_{n=1}^N p_{k,n} \le P, \forall t \in {\cal T},  \forall k \in {\cal K}$, where $p_{k,n} \ge 0 $ denotes the transmit power allocated to the $n$-th sub-carrier
%at user $k$ at time slot $n$.
Let ${\bm x}\triangleq  \left[x_{0},x_{1},\ldots, x_{N-1}  \right]^T$
denote the OFDM pilot symbol (without CP) sampled in the time domain.
%\vspace{-0.3cm}
%\begin{align}\label{OFDM_sym}
%X_{k,n}=\sqrt{\frac{P}{|{\cal J}_{k}|}} \delta_{k,n} S_{k,n} ,  \quad \forall t \in {\cal T},  \forall n \in {\cal N}, \forall k \in {\cal K}
%\end{align}
%where $S_{k,n}$ denotes the pilot symbol which is simply set as $S_{k,n}=1$ for ease of exposition, and we have $\left\|{\bm x}\right\|^2=P$.
%Before transmission, each OFDM symbol ${\bm x}$ is first transformed into the time domain via an $N$-point inverse DFT (IDFT), and then appended by a cyclic prefix (CP) of length $L_{cp}$ to mitigate the inter-symbol-interference (ISI), which is assumed to satisfy $L_{cp}\ge L$. 
Note that during the transmission of this OFDM symbol,
\rev{the effective channel ${\bm h}$ is made artificially time-varying by tuning the IRS training reflection coefficients ${\bm \theta}$ over different sampling periods within one OFDM symbol to facilitate the cascaded channel estimation}. Accordingly, the resultant ALTV channel at sampling period $n$, denoted by ${\bm h}^{(n)}$, is given by
%\begin{align}\label{superposed0}
%{\bm h}^{(n)}\triangleq \left[h_0^{(n)},  h_1^{(n)},\ldots,h_L^{(n)}\right]^T ={\bm Q} {\bm \theta}+{\bm d},
%\end{align}
\vspace{-0.1cm}
\begin{equation}\label{superposed2}
\resizebox{0.44\textwidth}{!}{$\hspace{-0.3cm}
\underbrace{\begin{bmatrix}
	h_0^{(n)}\\
	h_1^{(n)}  \\
	\vdots \\
	h_{L-1}^{(n)}
	\end{bmatrix}}_{{\bm h}^{(n)}}
=
\underbrace{\begin{bmatrix}
	d_1\\
	d_2  \\
	\vdots \\
	d_{L-1}
	\end{bmatrix}}_{{\bm d}}+
\underbrace{\begin{bmatrix}
	q_{1,0}      	&q_{2,0}        		&\cdots &q_{M,0}  \\
	q_{1,1}      	&q_{2,1}        		&\cdots &q_{M,1}  \\
	\vdots   		& \vdots  		&       &\vdots   		\\
	q_{1,L-1}      	&q_{2,L-1}        		&\cdots &q_{M,L-1}  \\   
	\end{bmatrix}}_{{\bm{{Q}}}}
\underbrace{\begin{bmatrix}
	\theta_1^{(n)}\\
	\theta_2^{(n)}  \\
	\vdots \\
	\theta_{M}^{(n)}
	\end{bmatrix}}_{{\bm \theta}^{(n)} }
\hspace{-0.2cm}$}
\end{equation}
where ${{\bm q}}_{m} = \left[q_{m,0},  q_{m,1},\ldots,q_{m,L-1}\right]^T$ and $\theta_m^{(n)}$ denotes the phase shift of sub-surface $m$ at sampling period $n$ with $m\in {\cal M}$ and $n\in \{0,1,\ldots, N-1\}$.
%After CP removal, the equivalent baseband received signal in the time domain is given by 
%%\vspace{-0.5cm}
%\begin{align}\label{rec_OFDM}
%\underbrace{\begin{bmatrix}
%	y_0\\
%	y_1  \\
%	\vdots \\
%	y_{N-1}
%	\end{bmatrix}}_{{\bm{{y}}}}
%=& \underbrace{\begin{bmatrix}
%	h_0^{(0)}      	&0        		&\cdots &0        		&h_{L-1}^{(0)}  &\cdots &h_1^{(0)}\\
%	h_1^{(1)}      	& h_0^{(1)}     &       &0        		&0         		&       &h_2^{(1)}\\
%	\vdots   		& \vdots  		&       &\vdots   		&\vdots    		&     	&\vdots\\
%	h_{L-1}^{(L-1)}  	&h_{L-2}^{(L-1)}  &       & 0       		&0          	&      	&0\\
%	0        		&h_{L-1}^{(L)}&       & 0       		&0         		&       &0\\
%	\vdots   		& \vdots  		&     	&\vdots   		&\vdots    		&       &\vdots\\
%	0        		&0        		& \cdots& h_{L-1}^{(N-1)} &h_{L-2}^{(N-1)}  &\cdots &h_0^{(N-1)}
%	\end{bmatrix}}_{{\bm{{H}}}}
%  \underbrace{\begin{bmatrix}
%	x_0\\
%	x_1  \\
%	\vdots \\
%	x_{N-1}
%	\end{bmatrix}}_{{\bm{{ x}}}}
%+\underbrace{\begin{bmatrix}
%	v_1\\
%	v_2  \\
%	\vdots \\
%	v_{N-1}
%	\end{bmatrix}}_{{\tilde\bm{{v}}}},
%\end{align}
%where 
%${\bm y}$ is the received OFDM symbol,
%$h_l^{(n)}$ denotes the effective CIR of the $l$-th tap at sampling period $n$ with $l\in \{0,1,\ldots, L-1\}$ and $n\in \{0,1,\ldots, N-1\}$,
%and ${\bm v} \sim {\mathcal N_c }({\bm 0}_{N\times1}, \sigma^2{\bm I}_N )$ is the additive white Gaussian noise (AWGN) vector at the BS with $\sigma^2$ being the noise power. 
After CP removal at the user, the equivalent baseband signal reflected by sub-surface $m$ in the time domain is given by \eqref{rec_sub-surface} at the top of this page, 
and the equivalent baseband signal through the direct link in the time domain is ${\bm{{r}}}_0={\bm{{X}}}{\bm d}$.
%\begin{align}\label{rec_direct}
%{\bm{{r}}}_0={\bm{{X}}}{\bm d}.
%\end{align}
Based on \eqref{rec_sub-surface}, the equivalent baseband received signal in the time domain can be rewritten as
\setcounter{equation}{8} 
\vspace{-0.2cm}
\begin{align}\label{rec_y}
\vspace{-0.2cm}
{\bm{{y}}}&=\underbrace{{\bm{{X}}}{\bm d}}_{{\bm{{r}}}_0}+ \sum_{m=1}^M \underbrace{{\bm{{ \Theta}}_m}{\bm{{X}}}{\bm q}_{m}}_{{\bm{{r}}}_m}+\tilde{\bm{{v}}} = {\bm{{\Xi}}}{\bm \lambda}+\tilde{\bm{{v}}}
\end{align}
where ${\bm{{\Xi}}}\triangleq\left[\bm{{ \Theta}}_0{\bm{{X}}},{\bm{{ \Theta}}_1}{\bm{{X}}},\ldots,{\bm{{ \Theta}}_M}{\bm{{X}}} \right]$ with $\bm{{ \Theta}}_0={\bm I}_N$,
${\bm \lambda} \triangleq \left[{\bm d}^T,{\bm q}_{1}^T,\ldots,{\bm q}_{M}^T\right]^T$,
and $\tilde{\bm{{v}}}\sim {\mathcal N_c }({\bm 0}, \sigma^2{\bm I}_{N} )$ is the AWGN vector at the user.
 \rev{It is noted that ${\bm \lambda} $ includes all the required CSI of the direct channel ${\bm d}$ and the cascaded channels $\{{\bm q}\}_{m=1}^M$.}
With ${\bm{{\Xi}}}^{\dagger}=\left( {\bm{{\Xi}}}^H{\bm{{\Xi}}} \right)^{-1}{\bm{{\Xi}}}^H$
denoting the left pseudo-inverse of ${\bm{{\Xi}}}$ and left-multiplying ${\bm{{y}}}$ in \eqref{rec_y} by ${\bm{{\Xi}}}^{\dagger}$,
we obtain the LS estimate of ${\bm \lambda}$ as
\begin{align}\label{LS_est}
\vspace{-0.2cm}
\hat{\bm \lambda}={\bm{{\Xi}}}^{\dagger} {\bm{{y}}}={\bm \lambda}+{\bm{{\Xi}}}^{\dagger}\tilde{\bm{{v}}}.
\end{align}
Note that for the LS
channel estimation in \eqref{LS_est}, $N \ge L(M+1)$ is the necessary condition to ensure the 
existence of ${\bm{{\Xi}}}^{\dagger}$, \rev{implying that the total number of sampling periods should be no less than that of channel
coefficients.}
Similarly, since the matrix inversion operation for computing ${\bm{{\Xi}}}^{\dagger}$ generally has a cubic time complexity in terms of $L(M+1)$,
it requires a proper design of ${\bm{{\Xi}}}$ to reduce such complexity as well as achieve the minimum MSE of channel estimation based on \eqref{LS_est}, as will be shown in the next.
%and may lead to suboptimal channel estimation due to the potential noise enhancement if ${\bm{{\Xi}}}$ is ill-conditioned, which thus requires proper design of ${\bm{{\Xi}}}$.
%the left pseudo-inverse of ${\bm{{\Xi}}}$ exists if and only if ${\bm{{\Xi}}}$ is of full column rank, which requires.
%This is the necessary but not necessarily sufficient condition for achieving the full column rank of ${\bm{{\Xi}}}$ and thus we need to carefully design ${\bm{{X}}}$ and $\left\{{\bm{{ \Theta}}_m}\right\}_{m=1}^M$ to ensure the existence of a full-row-rank ${\bm{{\Xi}}}$,
%which will be specified in the next section.

%\begin{figure*}[!t]
%	\centering
%	\includegraphics[width=7in]{pilot_sequence_V2.eps}
%	%\setlength{\abovecaptionskip}{-6pt}
%	\caption{An illustration of OFDM pilot sequence \rev{(will be removed)}.}
%	\label{OFDM pilot sequence}
%	\vspace{-0.4cm}
%\end{figure*}
\begin{table*}[!t]
	\begin{center}\caption{Comparison of Two Proposed Channel Estimation Schemes for IRS-Assisted OFDM}\label{Table of estimation}
		\vspace{0.3cm}
		\resizebox{0.8\textwidth}{!}{
			\begin{tabular}{|c|c|c|c|c|c|}
				\hline
				& \begin{tabular}[c]{@{}c@{}}Complexity\\ (number of multiplications)\end{tabular} & \begin{tabular}[c]{@{}c@{}}Training time\\ (sampling periods)\end{tabular} & Channel condition & \begin{tabular}[c]{@{}c@{}}Minimum\\ MSE\end{tabular} & Processing domain \\ \hline
				Scheme 1  & $N_0L(I_0+1) +L I_0(M+1)$                                                        & $(M+1)(N_0+L_{cp})$                                                        & Arbitrary channels    & $\frac{\sigma^2}{\gamma_1(M+1)}$                      & Frequency  \\ \hline
				Scheme 2 & $NL(M+1)$                                                                        & $N+L_{cp}$                                                            & Single-path IRS$\rightarrow $user channel ($L_2=1$)  & $\frac{\sigma^2}{\gamma_2 N}$                         & Time       \\ \hline
			\end{tabular}
		}
	\end{center}
	\vspace{-0.55cm}
\end{table*}
%\section{Optimal Training Design}
%From \eqref{LS_est}, the MSE of the channel estimation based on the sampling-wise varying IRS reflection is given by
The MSE of the LS channel estimation in \eqref{LS_est} is given by
\begin{align}\label{MSE0}
\vspace{-0.2cm}
\varepsilon_2&=\frac{1}{L(M+1)}  {\mathbb E}\left\{  \left\| \hat{\bm \lambda}
-{\bm \lambda}\right\|^{2}
\right\}
=\frac{1}{L(M+1)} {\mathbb E}\left\{ \left\|  {\bm{{\Xi}}}^{\dagger}\tilde{\bm{{v}}} \right\|^{2}\right\}
\notag \\
&=\frac{1}{L(M+1)}\text{tr} \left\{  {\bm{{\Xi}}}^{\dagger} {\mathbb E}\left\{ \tilde{\bm v}\tilde{\bm v}^H  \right\}     \left({\bm{{\Xi}}}^{\dagger}\right)^H   \right\}\notag \\
%=\frac{\sigma^2}{L(M+1)}\text{tr} \left\{  {\bm{{\Xi}}}^{\dagger}      \left({\bm{{\Xi}}}^{\dagger}\right)^H   \right\}
&=\frac{\sigma^2}{L(M+1)}\text{tr} \left\{  \left( {\bm{{\Xi}}}^H{\bm{{\Xi}}} \right)^{-1}   \right\}
\end{align}
where ${\mathbb E}\left\{ \tilde{\bm v}\tilde{\bm v}^H  \right\}=\sigma^2 {\bm I}_N$.
%Accounting for the constraints on the training design, the optimization problem for minimizing
%the MSE in \eqref{MSE0} is formulated as follows (with constant/irrelevant terms omitted for brevity).
%\begin{align}
%\text{(P1):}~
%& \underset{ \left\{\theta_m^{(n)}\right\},{\bm x}  }{\text{min}}
%& &\hspace{-3cm} \text{tr}\left\{  \left( {\bm{{\Xi}}}^H{\bm{{\Xi}}} \right)^{-1}   \right\} \label{obj_P1}\\
%& \text{s.t.} & &\hspace{-3cm} |\theta_m^{(n)}|=1, \forall  m\in {\cal M} , \forall t \in \{0,1,\ldots, N-1\}\label{con1_P1}\\
%& & & \hspace{-3cm}\left\|{\bm x}\right\|^2=P\label{con2_P1}.
%\end{align}
For the MSE in \eqref{MSE0}, we have 
\vspace{-0.2cm}
\begin{align}\label{corr1}
&{\bm{{\Xi}}^H}\hspace{-0.05cm}{\bm{{\Xi}}}\hspace{-0.1cm}=\hspace{-0.1cm}\left[{\bm{{ \Theta}}_0}\hspace{-0.05cm}{\bm{{X}}},{\bm{{ \Theta}}_1}\hspace{-0.05cm}{\bm{{X}}},\ldots,{\bm{{ \Theta}}_M}\hspace{-0.05cm}{\bm{{X}}} \right]^H\hspace{-0.13cm}
\left[{\bm{{ \Theta}}_0}\hspace{-0.05cm}{\bm{{X}}},{\bm{{ \Theta}}_1}\hspace{-0.05cm}{\bm{{X}}},\ldots,{\bm{{ \Theta}}_M}\hspace{-0.05cm}{\bm{{X}}} \right]\notag\\
&\stackrel{(b)}{=}\hspace{-0.1cm}\begin{bmatrix} 
\hspace{-0.08cm}{\bm{{X}}^H}\hspace{-0.08cm}{\bm{{X}}} & {\bm{{X}}^H}\hspace{-0.08cm} {\bm{{ \Theta}}_0^H}\hspace{-0.08cm}  {\bm{{ \Theta}}_1}\hspace{-0.08cm}{\bm{{X}}} & \cdots & {\bm{{X}}^H}\hspace{-0.08cm} {\bm{{ \Theta}}_0^H}\hspace{-0.08cm}{\bm{{ \Theta}}_M}\hspace{-0.08cm}{\bm{{X}}}\\ 
\hspace{-0.08cm}{\bm{{X}}^H}\hspace{-0.08cm} {\bm{{ \Theta}}_1^H}\hspace{-0.08cm} {\bm{{ \Theta}}_0} \hspace{-0.05cm}{\bm{{X}}} & {\bm{{X}}^H}\hspace{-0.08cm} {\bm{{X}}} & \cdots & {\bm{{X}}^H}\hspace{-0.08cm}{\bm{{ \Theta}}_1^H}\hspace{-0.08cm}{\bm{{ \Theta}}_M}\hspace{-0.08cm}{\bm{{X}}}\\ 
\hspace{-0.08cm} \vdots & \vdots &  \ddots & \vdots\\ 
\hspace{-0.08cm}{\bm{{X}}^H}\hspace{-0.08cm}{\bm{{ \Theta}}_M^H} \hspace{-0.08cm}{\bm{{ \Theta}}_0}\hspace{-0.05cm}{\bm{{X}}} & {\bm{{X}}^H}\hspace{-0.08cm}{\bm{{ \Theta}}_M^H}\hspace{-0.08cm}{\bm{{ \Theta}}_1}\hspace{-0.08cm}{\bm{{X}}} & \cdots & {\bm{{X}}^H} \hspace{-0.08cm}{\bm{{X}}}\hspace{-0.15cm}
\end{bmatrix}\hspace{-0.1cm}
\end{align}
where $(b)$ holds since ${\bm{{ \Theta}}_m^H}{\bm{{ \Theta}}_m}={\bm I}_N, \forall m \in {\cal M}$.
%due to the unit-modulus constraint of $\theta_m^{(n)}, \forall m \in {\cal M},\forall t\in \{0,1,\ldots, N-1\}$, we have ${\bm{{ \Theta}}_m^H}{\bm{{ \Theta}}_m}={\bm I}_N, \forall m \in {\cal M}$, and thus \eqref{corr1} can be simplified as
%\begin{align}
%{\bm{{\Xi}}^H}{\bm{{\Xi}}}&=\begin{bmatrix} 
%{\bm{{X}}^H}{\bm{{X}}} & {\bm{{X}}^H} {\bm{{ \Theta}}_1}{\bm{{X}}} & \cdots & {\bm{{X}}^H} {\bm{{ \Theta}}_M}{\bm{{X}}}\\ 
%{\bm{{X}}^H} {\bm{{ \Theta}}_1^H} {\bm{{X}}} & {\bm{{X}}^H} {\bm{{X}}} & \cdots & {\bm{{X}}^H}{\bm{{ \Theta}}_1^H}{\bm{{ \Theta}}_M}{\bm{{X}}}\\ 
%\vdots & \vdots &  \ddots & \vdots\\ 
%{\bm{{X}}^H}{\bm{{ \Theta}}_M^H} {\bm{{X}}} & {\bm{{X}}^H}{\bm{{ \Theta}}_M^H}{\bm{{ \Theta}}_1}{\bm{{X}}} & \cdots & {\bm{{X}}^H} {\bm{{X}}}
%\end{bmatrix}.
%\end{align} 
\rev{In particular, the OFDM pilot symbol ${\bm{{x}}}$ and the IRS training reflection coefficients $\left\{\theta_m^{(n)}\right\}$ should be jointly designed to achieve
${\bm \Xi}  {\bm \Xi^H} =c{\bm I}_{L(M+1)}$ for the MSE minimization in \eqref{MSE0},} which is equivalent to the following conditions:
%, which requires the careful design of ${\bm{{X}}}$ and $\left\{{\bm{{ \Theta}}_m}\right\}_{m=1}^M$. 
%To achieve this, we need to carefully design both ${\bm{{X}}}$ and $\left\{{\bm{{ \Theta}}_m}\right\}_{m=1}^M$ to satisfy
 \begin{align}
 \vspace{-0.3cm}
 {\bm{{X}}^H}{\bm{{X}}}&=c{\bm I}_{L}\label{X_relate}\\
 {\bm{{X}}^H}{\bm{{ \Theta}}_m^H}{\bm{{ \Theta}}_{m'}}{\bm{{X}}}&={\bm 0}_{L\times L}\label{IRS_con}
 \end{align}
 where $m\ne m', \forall m,m'\in \{0,1,\ldots, M\}$ and $c$ is a positive constant to be determined later.

 Let ${\bm{C}}\triangleq \left[{\bm{{x}}}_{0}, {\bm{{x}}}_{-1}, \ldots, {\bm{{x}}}_{-N+1}\right]$ denote the circulant matrix
 	generated from the OFDM pilot symbol ${\bm{{x}}}$, where ${\bm{{x}}}_{-n}$ is the circularly shifted version of ${\bm{{x}}}$ by $n$ steps in the downward direction with $n=0,1,\ldots N-1$.
 As shown in \eqref{rec_sub-surface}, ${\bm{{X}}}$ is an $N \times L$ matrix consisting of the first $L$ columns of the circulant matrix
 ${\bm{C}}$, i.e., ${\bm{{X}}}=\left[{\bm{{x}}}_{0}, {\bm{{x}}}_{-1}, \ldots, {\bm{{x}}}_{-L+1}\right]$.
%As shown in \eqref{rec_sub-surface}, ${\bm{{X}}}$ is a Toeplitz matrix generated from the OFDM pilot symbol ${\bm{{x}}}$, i.e., each column of ${\bm{{X}}}$ is a cyclically shifted version of ${\bm{{x}}}$. 
\rev{The condition in \eqref{X_relate} indicates that any two columns in ${\bm{{X}}}$ are orthogonal, i.e., ${\bm{{x}}}_{-l}^H{\bm{{x}}}_{-l'}=0$ with $l\ne l', \forall l,l'\in \{0,1,\ldots, L-1\}$, which requires that the auto-correlation of ${\bm{{x}}}$ should be zero.} This can be achieved by setting the OFDM pilot symbol ${\bm{{x}}}$ as a Zadoff-Chu sequence \cite{Polyphase1972Polyphase} with each element given by
\begin{align}\label{Zadoff-Chu}
\vspace{-0.2cm}
x_{n}=\gamma_2 \cdot e^{-j \frac{\pi \omega n^2}{N}}, \quad n=0,1,\ldots N-1
\end{align}
where $\omega$ is an integer relatively prime to $N$ and $\gamma_2$ denotes the average power at each sampling period. \rev{Note that as compared to the design of ${\bm{{x}}}$ to achieve \eqref{X_relate},
it is much more challenging to jointly design the IRS training reflection coefficients $\left\{\theta_m^{(n)}\right\}$ to satisfy the condition in \eqref{IRS_con}. Fortunately, we notice that given ${\bm{{x}}}$ in \eqref{Zadoff-Chu}, the $N$ cyclically shifted versions of ${\bm{{x}}}$ in ${\bm{C}}$ are pairwise orthogonal to each other; however, 
only $L$ of them are involved in ${\bm{{X}}}$ (see \eqref{rec_sub-surface}), while the remaining $N-L $ cyclically shifted versions of ${\bm{{x}}}$ have not been exploited yet.}
%As such, by denoting ${\bar{\bm{{X}}}}_m\triangleq{\bm{{ \Theta}}_{m}}{\bm{{X}}}$, we can assign the remaining $N-L$ cyclically shifted versions of ${\bm{{x}}}$ to each $${\bar{\bm{{X}}}}_m$
%Inspired by this, we design the IRS reflection coefficients $\left\{\theta_m^{(n)}\right\}$ to achieve the orthogonality between each ${\bar{\bm{{X}}}}_m\triangleq{\bm{{ \Theta}}_{m}}{\bm{{X}}}$, i.e., ${\bar{\bm{{X}}}}_m^H {\bar{\bm{{X}}}}_{m'}={\bm 0}_{L\times L}$ with $m\ne m', \forall m,m'\in \{0,1,\ldots, M\}$.
Inspired by this, we let ${\bar{\bm{{X}}}}_m\triangleq{\bm{{ \Theta}}_{m}}{\bm{{X}}}$ and disjointly assign the remaining $N-L $ cyclically shifted versions of ${\bm{{x}}}$, i.e., $\left\{{\bm{{x}}}_{-n}\right\}_{n=L}^{N-1}$ for each ${\bar{\bm{{X}}}}_m$ to achieve pairwise orthogonality, i.e., ${\bar{\bm{{X}}}}_m^H {\bar{\bm{{X}}}}_{m'}={\bm 0}_{L\times L}$ with $m\ne m', \forall m,m'\in \{0,1,\ldots, M\}$.
For example, we can set ${\bar{\bm{{X}}}}_m=\left[{\bm{{x}}}_{-mL}, {\bm{{x}}}_{-mL-1}, \ldots, {\bm{{x}}}_{-(m+1)L+1}\right]$ with $m \in \{0,1,\ldots, M\}$,
%Accordingly, ${\bar{\bm{{X}}}}_m$ can be designed as the matrix that circularly shifts each column of ${\bm{{X}}}$ by $mL$ steps in the downward direction \rev{(as shown in Fig.~\ref{OFDM pilot sequence}, which will be removed due to the limited space)}, 
and thus have ${\bm{{ \Theta}}_{m}}={\rm diag}({\bm{{x}}}_{-mL-l})\left({\rm diag} ({\bm{{x}}}_{-l})\right)^{-1}, \forall l \in \{0,1,\ldots, L-1\}$, in which the corresponding IRS training reflection coefficients $\left\{\theta_m^{(n)}\right\}$ are given by
%{\bar{\bm{{X}}}}_m{\bm{{X}}}^{-1}=
%as an $mL$-step downward cyclic-shift each column of ${\bm{{X}}}$
%circularly shifts the elements in array A by K positions.
%circularly shifts the values in array A by K positions along dimension dim.
 \begin{align}\label{IRS_coeff}
 \vspace{-0.2cm}
 \theta_m^{(n)}&=\frac{x_{n-mL}}{x_{n}}=\frac{e^{-j \frac{\pi \omega (n-mL)^2}{N}}}{e^{-j \frac{\pi \omega n^2}{N}}}=e^{j \frac{\pi \omega (2n-mL)mL}{N}}
 \end{align}
 with $m\in {\cal M},n\in \{0,1,\ldots, N-1\}$.
%To achieve \eqref{X_relate}, we can consider the Zadoff-Chu sequences as the OFDM pilot symbol ${\bm{{x}}}$ in the time domain, since the auto correlation of a Zadoff-Chu sequence \cite{Polyphase1972Polyphase} with a cyclically shifted version of itself is zero, i.e., it is non-zero only at one instant which corresponds to the cyclic shift.
Based on \eqref{Zadoff-Chu} and \eqref{IRS_coeff}, we obtain $c=\gamma_2 N$ and thus the minimum MSE of \eqref{MSE0} is given by $\varepsilon_2^{\min}=\frac{\sigma^2}{\gamma_2 N}$.
Moreover, we have ${\bm{{\Xi}}}^{\dagger}=\frac{1}{\gamma_2 N}{\bm{{\Xi}}}^H$, which avoids
the matrix inversion operation of cubic-order complexity.

\begin{figure}[!t]
	\centering
	\includegraphics[width=3.5in]{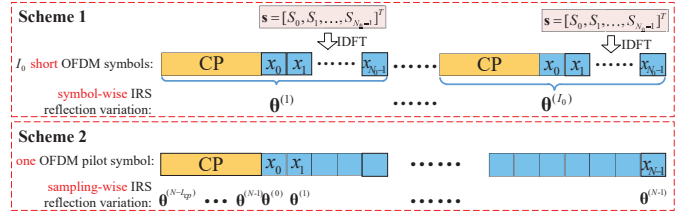}
	\caption{\rev{Illustration of OFDM symbol structures and IRS reflection variations for the two proposed channel estimation schemes.}}
	\label{OFDM_structure}
	\vspace{-0.5cm}
\end{figure}

\rev{Finally, we illustrate the OFDM symbol structures (short vs. long) and IRS reflection variations (symbol-wise vs. sampling-wise)} for the two proposed channel estimation schemes in Fig.~\ref{OFDM_structure} and summarize their comparison in Table~\ref{Table of estimation}.
%Both proposed algorithms achieve near-optimal performances with complexity linear in terms of the antenna size.
%leading to linear complexity with respect to the number of transmit antennas.
\rev{Note that owing to the perfect orthogonality of the joint training design for the pilot signal and IRS reflection pattern,
each proposed channel estimation scheme achieves its minimum MSE with low complexity (linear with respect to $M$ and/or $L$).}
Generally speaking, Scheme~1 requires lower channel estimation complexity than Scheme~2, but incurs more training time as well as higher minimum MSE (as will be shown by simulations in Section~\ref{Sim}) for the case of $L_2=1$. 
%In particular, as $N_0\ge L$ and $L_{cp}\ge L$, the minimum training overhead of Scheme 1 is nearly twice that of Scheme 2.
Furthermore, the MSE gain of~Scheme~2 over Scheme 1 in dB for the case of $L_2=1$ is given by
 \begin{align}\label{MSE_gain}
 G=-10\log_{10} \left(\frac{\varepsilon_2^{\min}}{\varepsilon_1^{\min}}\right)=10\log_{10}
 \frac{\gamma_2 N}{\gamma_1(M+1)}.
\end{align}
\rev{The MSE gain in \eqref{MSE_gain} is due to two factors: one is the average power ratio (which is associated with the training time ratio given the same total power budget, i.e., $\gamma_2/\gamma_1=\eta_1/\eta_2$); the other is the IRS reflection variation ratio (sampling-wise vs. symbol-wise) between the two channel estimation schemes, i.e., $N/(M+1)$.}

\vspace{-0.2cm}
\section{Simulation Results}\label{Sim}
In this section, we present simulation results to
validate the effectiveness of our proposed channel estimation schemes.
The IRS consists of $M_0=15\times 9=135$ reflecting elements with half-wavelength spacing, which are divided into $M=15$ sub-surfaces, each with $\mu= {M}_0/M=9$ elements.
The maximum delay spreads of both the direct BS$\rightarrow$user channel and the cascaded BS$\rightarrow$IRS$\rightarrow$user channel are set as $L_r=L_d=L=8$, while the exact settings of $L_1$ and $L_2$ for the BS$\rightarrow$IRS and IRS$\rightarrow$user channels will be specified later depending on the scenarios.
For the BS$\rightarrow$user and BS$\rightarrow$IRS links, their frequency-selective Rayleigh fading channels are characterized by an exponentially decaying power delay profile with a root-mean-square delay spread and a spread power decaying factor of $2$.
%, where each tap is generated according to Rayleigh distribution and the spread power decaying factor is set to $2$.
For the IRS$\rightarrow$user link modeled by the frequency-selective Rician fading channel (i.e., $L_2\ge 1$ in general), the first tap is set as the LoS component and the remaining taps are Non-LoS (NLoS) Rayleigh fading  components, with $\kappa$ being the Rician factor.
The distance-dependent channel path loss is modeled as $\gamma=\gamma_0/ D^\alpha$, where $\gamma_0$ is the reference path loss at a distance of 1 meter (m), $D$ is the individual link distance, and $\alpha$ is the path loss exponent.
Moreover, the distance between the BS and IRS is $50$ m and the user is located in the proximity of the IRS with a distance of $1.5$ m.

For Scheme 1, since $N_0\ge L$ and $L_{cp}\ge L$, it requires at least $\eta_{1,\min}=2L(M+1)$ sampling periods for channel training;
while for Scheme 2, by letting $N=L(M+1)$ and $L_{cp}=L$,
the minimum training time in terms of sampling periods is $\eta_{2,\min}=L(M+1)+L=L(M+2)$.
\rev{Thus, Scheme~2 only requires about half of the training time of Scheme 1.
%Thus, $\eta_{1,\min}$ is about twice of $\eta_{2,\min}$. 
For fair comparison, we consider the same total power budget $P$ for the channel training such that we have $\gamma_1=\frac{P}{\eta_{1,\min}}$ and $\gamma_2=\frac{P}{\eta_{2,\min}}$ for the two proposed schemes, respectively.}
The SNR is defined as the ratio between the average power of the received signal at each sampling period and the noise power at the user, which is given by
\begin{equation}
\resizebox{0.48\textwidth}{!}{$
\text{SNR}= {\mathbb E}\left\{\frac{P \left\|{\bm d}+{\bm Q} {\bm \theta}\right\|^2  }{\sigma^2(N+L_{cp})} \right\}= \frac{P( M_0\gamma_0^2 D_1^{-\alpha_1}D_2^{-\alpha_2}+\gamma_0D_3^{-\alpha_3} )   }{\sigma^2(N+L_{cp})}
$}\notag
\end{equation}
where $D_1$, $D_2$, and $D_3$ denote the distances of the IRS$\rightarrow$user, BS$\rightarrow$IRS, and (direct) BS$\rightarrow$user links, respectively; $\alpha_1$, $\alpha_2$, and $\alpha_3$ denote the path loss exponents of these links, which are set as $2.2$, $2.4$, and $3.6$, respectively; the path loss at the reference distance is set as $\gamma_0=-30$~dB for each individual link; and the noise power is set as $\sigma^2=-80$~dBm.
We calculate the normalized MSE (with respect to the overall channel gain, i.e., ${ \left\|\left[{\bm d}, {\bm Q} \right]\right\|^{2}_F}$ or ${ \left\|{\bm \lambda}\right\|^{2}}$) over $10000$ independent fading channel realizations.
%, which is given by
%\begin{align}\label{normalizedMSE}
%{\bar\varepsilon}=\frac{1}{L(M+1)}    {\mathbb E}\left\{\left\|\left[{\hat{\bm d}}, {\hat{\bm Q}} \right]
%-\left[{\bm d}, {\bm Q} \right]\right\|^{2}_F
%\Big/{ \left\|\left[{\bm d}, {\bm Q} \right]\right\|^{2}_F} \right\} 
%=\frac{1}{L(M+1)}    {\mathbb E}\left\{\left\| \hat{\bm \lambda}
%-{\bm \lambda}\right\|^{2}
%\Big/{ \left\|{\bm \lambda}\right\|^{2}} \right\} .
%\end{align}

\begin{figure}[!t]
	\centering
	\includegraphics[width=3.2in]{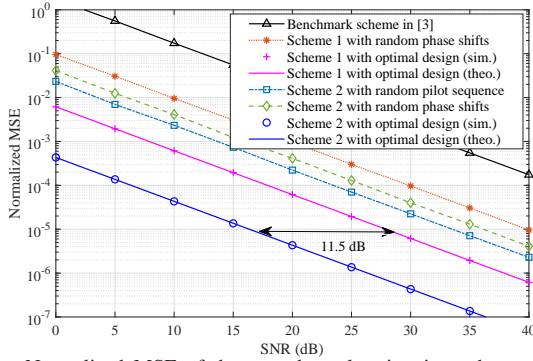}
	\caption{Normalized MSE of the two channel estimation schemes vs. SNR with $L_1=8$ and $L_2=1$.}
	\label{Fast_IRS_comp}
	\vspace{-0.5cm}
\end{figure}
In Fig.~\ref{Fast_IRS_comp}, we compare the normalized MSE of the two proposed channel estimation schemes with $L_1=8$ and $L_2=1$ (i.e., the IRS$\rightarrow$user link with the LoS component only).
It is observed that the theoretical analysis of MSE given in \eqref{MSE_cp} and \eqref{MSE0} is in perfect agreement with the simulation results.
\rev{Moreover, Scheme 2 achieves up to $11.5$~dB SNR gain over Scheme 1 (albeit Scheme 1 spends longer time for channel training), which corroborates the analytical MSE gain $ G=10\log_{10}
\frac{\gamma_2 N}{\gamma_1(M+1)}=11.53$~dB given in \eqref{MSE_gain}.}
%, where the contributions of this MSE gain have been explained at the end of Section~\ref{Sampling}.
%\rev{Finally, compared to the benchmark schemes (OF/OFF and random reflection patterns) requiring cubic time complexity due to the inversion operation, the two proposed channel estimation schemes achieve the minimum MSE at a much lower complexity by adopting the joint orthogonal design for pilot sequences and IRS reflections.} 
Finally, we consider the benchmark designs where either the IRS training reflection coefficients or the pilot symbols are generated
with random phase shifts following the uniform distribution within $[0, 2\pi)$ for comparison.\footnote{The DFT-based IRS training reflection pattern for Scheme~1 is inapplicable to Scheme~2 since the resultant ${\bm{{\Xi}}}$ with the Zadoff-Chu pilot sequence is not of full rank.}
It is observed that the two proposed channel estimation schemes with optimal training designs significantly outperform their corresponding benchmark schemes.
%These results also verify the effectiveness of the proposed joint design of the pilot sequence and IRS reflection pattern with \emph{perfect orthogonality}.

\begin{figure}[!t]
	\centering
	\includegraphics[width=3.2in]{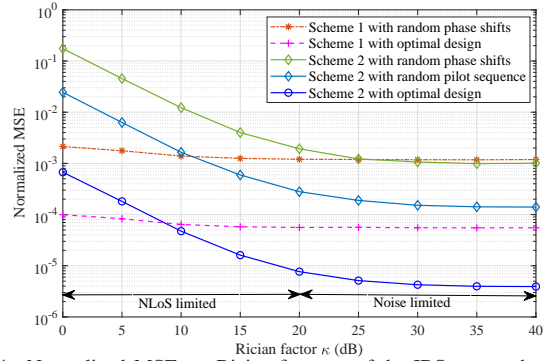}
	\caption{Normalized MSE vs. Rician factor $\kappa$ of the IRS$\rightarrow$user channel with SNR $=20$~dB, $L_1=7$, and $L_2=2$.}
	\label{Fast_IRS_comp_NLoS}
	\vspace{-0.5cm}
\end{figure}

In Fig.~\ref{Fast_IRS_comp_NLoS}, we examine the effect of NLoS interference in the IRS$\rightarrow$user link on the channel estimation performance for Scheme 2 as compared to Scheme 1, by showing the normalized MSE vs. this channel Rician factor $\kappa$ (dB) with SNR $=20$ dB, $L_1=7$, and $L_2=2$.
%In this case, the channel estimation performance of Scheme 2 is affected by both the multi-path interference and AWGN.
As the Rician factor $\kappa$ increases, it is observed that the normalized MSE of Scheme~2 decreases drastically in the range of $\kappa \in [0, 20]$~dB, while it approaches an error floor in the range of $\kappa \in [20, 40]$~dB.
This is due to the fact that given SNR $=20$ dB, the channel estimation error is mainly attributed to the NLoS interference when its power is higher than the noise power (i.e., $\kappa<20$~dB).
%Besides, we observe that for the SeUCE scheme, the proposed pilot tone allocation design always outperforms the permuted pilot tone allocation benchmark, regardless of the NLoS-limited or noise-limited region.
In contrast, we observe that the performance of Scheme 1 is almost unaffected by the NLoS component power, since it is applicable to arbitrary frequency-selective fading channels.
%It is worth pointing out that the normalized MSE of Scheme 1 slightly decreases as the increase of the Rician factor $\kappa$.
%This can be explained by the fact that although the MSE of Scheme 1 is determined by the noise power only and unaffected by the Rician factor $\kappa$, the overall channel gain (i.e., ${ \left\|\left[{\bm d}, {\bm Q} \right]\right\|^{2}_F}$ in \eqref{normalizedMSE}) slightly increases as the increase of the Rician factor $\kappa$, thus slightly decreasing the normalized MSE of Scheme 1.
\vspace{-0.1cm}
\section{Conclusions}
In this letter, we proposed two efficient channel estimation schemes for different
channel setups in the IRS-assisted OFDM system.
\rev{By exploiting the novel concept of sampling-wise varying IRS training reflection,} Scheme 2 was shown to achieve much lower MSE
with even less training time as compared to Scheme 1 under the condition of LoS-dominant IRS-user channel, but at the expense of slightly higher complexity.
Both proposed schemes were shown to achieve their respective minimum MSE via jointly optimized IRS training reflection pattern and pilot signal design. 
%\vspace{-0.4cm}

%Under the unit-modulus constraint, we have designed a novel reflection pattern for channel estimation and optimized the reflection coefficients with a low-complexity SCM method.
%Simulation results have verified the superior performance of our proposed methods over the existing schemes.  

\ifCLASSOPTIONcaptionsoff
  \newpage
\fi
\vspace{-0.1cm}
\bibliographystyle{IEEEtran}
% argument is your BibTeX string definitions and bibliography database(s)
\bibliography{IRS_OFDM}

% Generated by IEEEtran.bst, version: 1.13 (2008/09/30)
\begin{thebibliography}{10}
\providecommand{\url}[1]{#1}
\csname url@samestyle\endcsname
\providecommand{\newblock}{\relax}
\providecommand{\bibinfo}[2]{#2}
\providecommand{\BIBentrySTDinterwordspacing}{\spaceskip=0pt\relax}
\providecommand{\BIBentryALTinterwordstretchfactor}{4}
\providecommand{\BIBentryALTinterwordspacing}{\spaceskip=\fontdimen2\font plus
\BIBentryALTinterwordstretchfactor\fontdimen3\font minus
  \fontdimen4\font\relax}
\providecommand{\BIBforeignlanguage}[2]{{%
\expandafter\ifx\csname l@#1\endcsname\relax
\typeout{** WARNING: IEEEtran.bst: No hyphenation pattern has been}%
\typeout{** loaded for the language `#1'. Using the pattern for}%
\typeout{** the default language instead.}%
\else
\language=\csname l@#1\endcsname
\fi
#2}}
\providecommand{\BIBdecl}{\relax}
\BIBdecl

\bibitem{qingqing2019towards}
Q.~Wu and R.~Zhang, ``Towards smart and reconfigurable environment: Intelligent
  reflecting surface aided wireless network,'' \emph{IEEE Commun. Mag.},
  vol.~58, no.~1, pp. 106--112, Jan. 2020.

\bibitem{Huang2019Reconfigurable}
C.~{Huang}, A.~{Zappone}, G.~C. {Alexandropoulos}, M.~{Debbah}, and C.~{Yuen},
  ``Reconfigurable intelligent surfaces for energy efficiency in wireless
  communication,'' \emph{IEEE Trans. Wireless Commun.}, vol.~18, no.~8, pp.
  4157--4170, Aug. 2019.

\bibitem{yang2019intelligent}
Y.~Yang, B.~Zheng, S.~Zhang, and R.~Zhang, ``Intelligent reflecting surface
  meets {OFDM}: Protocol design and rate maximization,'' \emph{IEEE Trans.
  Commun.}, vol.~68, no.~7, pp. 4522--4535, Jul. 2020.

\bibitem{zheng2019intelligent}
B.~Zheng and R.~Zhang, ``Intelligent reflecting surface-enhanced {OFDM}:
  Channel estimation and reflection optimization,'' \emph{IEEE Wireless Commun.
  Lett.}, vol.~9, no.~4, pp. 518--522, Apr. 2020.

\bibitem{zheng2020intelligent}
B.~Zheng, C.~You, and R.~Zhang, ``Intelligent reflecting surface assisted
  multi-user {OFDMA}: Channel estimation and training design,'' \emph{arXiv
  preprint arXiv:2003.00648}, 2020.

\bibitem{jensen2019optimal}
T.~L. Jensen and E.~De~Carvalho, ``An optimal channel estimation scheme for
  intelligent reflecting surfaces based on a minimum variance unbiased
  estimator,'' in \emph{Proc. IEEE Int. Conf. Acoust., Speech, Signal Process.
  (ICASSP)}, Barcelona, Spain, May 2020, pp. 5000--5004.

\bibitem{chen2019channel}
J.~Chen, Y.-C. Liang, H.~V. Cheng, and W.~Yu, ``Channel estimation for
  reconfigurable intelligent surface aided multi-user {MIMO} systems,''
  \emph{arXiv preprint arXiv:1912.03619}, 2019.

\bibitem{you2019progressive}
\BIBentryALTinterwordspacing
C.~You, B.~Zheng, and R.~Zhang, ``Channel estimation and passive beamforming
  for intelligent reflecting surface: Discrete phase shift and progressive
  refinement,'' \emph{to appear in IEEE J. Sel. Areas Commun}, 2019. [Online].
  Available: \url{http://arxiv.org/abs/1912.10646}
\BIBentrySTDinterwordspacing

\bibitem{kay1993fundamentals}
S.~M. Kay, \emph{Fundamentals of statistical signal processing}.\hskip 1em plus
  0.5em minus 0.4em\relax Prentice Hall PTR, 1993.

\bibitem{Polyphase1972Polyphase}
D.~{Chu}, ``Polyphase codes with good periodic correlation properties,''
  \emph{IEEE Trans. Inf. Theory}, vol.~18, no.~4, pp. 531--532, Jul. 1972.

\end{thebibliography}

\end{document}